# PDM (Probes Distance Meter), Distance Measuring Instrument by Using Ultrasonic Propagation between Probes

Ridlo Qomarrullah, Abdurrachman Mappuji, Wahyu Apriliyanto, Ahmad Harist Julianto, Farahiyah Syarafina, Iswandi[1, a)]

[1]Universitas Gadjah Mada, Yogyakarta.

[a)]Corresponding author: ridlo.qomarrullah@mail.ugm.ac.id

**Abstract.** The innovation process is always required to produce more advanced technology and increasing productivity. PDM (Probes Distance Meter) comes as a technological innovation in measurement, PDM is a digital distance measuring instrument uses a pair of probes. PDM consists of ultrasonic transducers, a microcontroller, and screen which shows the measurement result. PDM utilizes ultrasonic wave principle by using two probes as reference points. A microcontroller is used for performing calculation that uses the distance equation. Innovation and Excellence of PDM are simple, ergonomic, modern, and able to show the measurement result automatically. PDM has additional features those are the storage media to save the measurement result with a micro-SD card, sound generation, and braille for the blinds. Those tools are expected to be suitable for common people, academics, or technicians to perform the measurement. The results from the laboratory testing show that the prototype of PDM has resolution of 1 cm, full scale of 200 cm, 11.1% repeatability, 77.6% accuracy, and a nonlinearity parameter of 0.16%. PDM is in development and it is potential to be a model of digital measuring instrument in the future that is simple, fast, and its additional features are capable to be used by the blind.

## INTRODUCTION

Measurement is a very important thing in daily life. The simplicity in the measurement highly depends on the measuring instrument utilization. The demand of modern measuring instrument is increasing either for simple or for complex measurement. The innovation in measurement field is always required to produce more advanced measuring instrument for improving ease of use and increasing productivity.

PDM (Probes Distance Meter) is an innovation of digital distance measuring instrument uses a pair of probes as reference of measurement. The main components of PDM are a microcontroller, ultrasonic transducers, and a pair of probes. PDM working mechanism is by pushing power button then the first probe transmits ultrasonic waves to the second probe automatically. After that, the microcontroller counts the time interval of propagated waves. Thus the distance traveled by the ultrasonic waves from the first probe to the second probe can be calculated by using math formula. The distance traveled by ultrasonic waves is the measurement result. It is displayed on the LCD in real time and also the result can be saved in the storage device that is MicroSD.

## MATERIALS AND METHOD

PDM consists of three main components those are a Arduino microcontroller, a pair of probes with ultrasonic transducer, and a LCD Display. The additional components of PDM are a speaker and a MicroSD card. Arduino in this instrument has three main function those are as ultrasonic generator, digital timer, and components controller.

As ultrasonic generator Arduino generates electric signal to be converted to ultrasonic waves by a transducer. The ultrasonic waves then propagated from the first probe to the second probe.

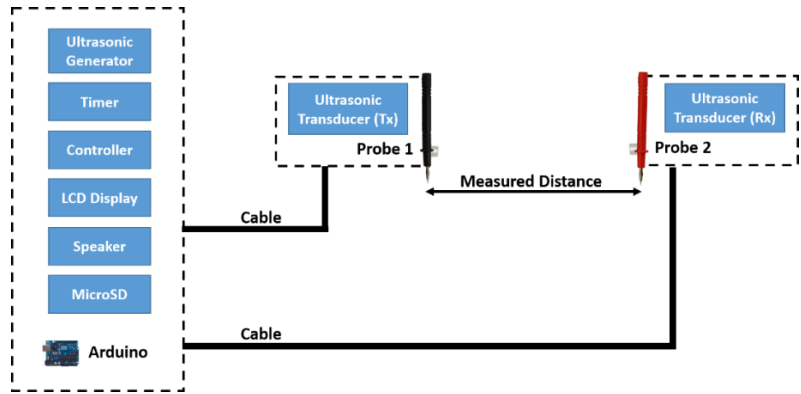

**FIGURE 1.** PDM working mechanism diagram

**TABLE 1.** PDM components specification

| Name | Specification | Amount |
|---|---|---|
| Microcontroller | Arduino Uno 8 MHz | 1 set |
| SRF05 | Working Frequency 40 kHz | 1 set |
| Ultrasonic Transducers | Working Frequency 40 kHz | 1 set |
| Jumper Cable | Copper | 4x5 meter |
| Battery | Li-Po 3 cell | 1 |
| MicroSD | 8 GB | 2 |
| LCD Display | 16x2 | 1 |
| Speaker | Stereo 5 V DC | 1 |

As digital timer, Arduino counts the time interval of ultrasonic waves propagation from the first probe to the second probe. Then Arduino performs the distance calculation by using the following distance formula:

$$S = v \times t \quad (1)$$

Where, $S$ = measured distance
$v$ = ultrasonic waves speed (348.61 m/s)
$t$ = propagation time

The instrument uses C program in Arduino to manage signal generation, distance calculation, measurement result, and display. The flowchart of the program is shown in Figure 2.

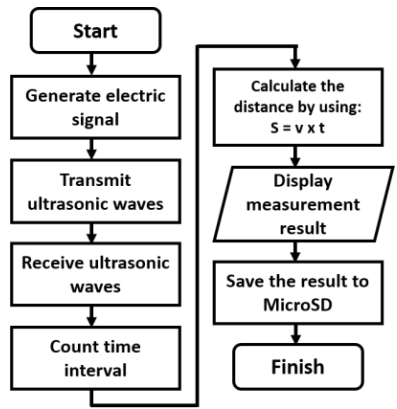

**FIGURE 2.** Programming flowchart

PDM uses ultrasonic waves with frequency of 40 kHz. The speed of ultrasonic waves depends on its medium. The theoretical expression for the speed of sound C in an ideal gas is (Bohn 1988):

$$C = \sqrt{\frac{\gamma P}{\rho}} \qquad (2)$$

Where, $C$ = speed of sound or ultrasonic waves
$\gamma$ = ratio of the specific heat of gas at constant pressure to that at constant volume
$P$ = ambient pressure
$\rho$ = gas density

The velocity of sound $C$ in dry air at 0°C and 1 atm (760 mm Hg) with 0.03 mol-% of carbon dioxide has the following experimentally verified values:

$$C = 331.45 \pm 0.05 \, m/s \qquad (3)$$

According to Eq. 2, then the speed of sound that depends on temperature is:

$$C = 331.45\sqrt{1 + \frac{t}{273}} \qquad (4)$$

Where, $t$ = temperature (Celcius)
$C$ = speed of sound or ultrasonic waves (m/s)

Thus in PDM experiment, in 29°C temperature the ultrasonic waves propagate with speed around of 348.61 m/s. PDM uses SRF05 ultrasonic transducers as transmitter and receiver with the specifications are shown in Table 2.

**TABLE 2.** SRF05 specification (Nexus Cyber Electronics 2015)

| SRF05 Parameter | Specification |
|---|---|
| range | 3 cm - 400 cm |
| interface | 1 pin I/O or 1 pin input & 1 pin output (adjustable) |
| output | digital pulse |
| ultrasonic burst frequency | 40 kHz |
| power supply | 5 V DC |

## RESULTS AND DISCUSSION

### Working Range and Resolution

PDM prototype is able to work on a range of 3 cm – 200 cm nowadays, but in the next development the maximum range is able to increase teoritically until 4,000 cm or 40 meters by upgrading to wireless probes. In many measuring instruments, it is important to know about the measurement resolution. Resolution is a parameter that shows the minimum input (in this case is the distance between probes) which is able to change the output. Considering the instrument is still in the prototype stage and the nature of ultrasonic waves, although the true resolution of the transducer is 3 mm, by using programming we round it off to be 1 cm resolution instead, because it is not standard to count the result in multiple of three like 3, 6, 9, and so on. For example, we give 3.2 cm distance for testing, the instrument output will show 3 cm. This resolution can be increased by upgrading the quality of ultrasonic transducers. Ultrasonic transducer with resolution of 1 mm is actually available with worth price.

### Accuracy and Repeatability of Instrument

Accuracy and precision (repeatability) are two of the most important parameter in measurement field. Accuracy is usually represented by inaccuracy. Inaccuracy is measured as a highest deviation of a value that is represented by the instrument from the ideal or true value of a stimulus at its input. The true value is attributed to the object of measurement and accepted as having a specified uncertainty because the true value is never absolutely sure (Fraden 2010). From laboratory test by comparing the measurement result to standard ruler as shown in the Table 3, the prototype of PDM has accuracy of 77.6%. It is not surprising that the accuracy is still below 90% because the nature of the ultrasonic transducer is sensitive to change in direction. A small vibration caused by the user might produce a random error. It is already decreased by putting flat circular seat for the probes, but random error is still high. In the next development, random error is able to be decreased by utilizing a low power laser and photodiode in first and second probe respectively to keep the ultrasonic probes facing each other in one line of direction.

**TABLE 3.** Measurement results of PDM compared to standard ruler

|  |  | Test Input [cm] (standard ruler) | | | |  | Test Input [cm] (standard ruler) | | | |
|---|---|---|---|---|---|---|---|---|---|---|
|  |  | 5 | 10 | 20 | 30 |  | 5 | 10 | 20 | 30 |
| Test Ouput [cm] | 1st | 5 | 11 | 21 | 31 | 11th | 5 | 10 | 20 | 30 |
|  | 2nd | 5 | 10 | 20 | 30 | 12nd | 5 | 10 | 20 | 30 |
|  | 3rd | 5 | 10 | 20 | 30 | 13rd | 5 | 10 | 20 | 30 |
|  | 4th | 5 | 10 | 20 | 30 | 14th | 5 | 10 | 20 | 30 |
|  | 5th | 5 | 10 | 20 | 30 | 15th | 5 | 10 | 20 | 30 |
|  | 6th | 5 | 10 | 20 | 30 | 16th | 5 | 10 | 20 | 30 |
|  | 7th | 5 | 10 | 20 | 30 | 17th | 5 | 10 | 20 | 30 |
|  | 8th | 5 | 10 | 20 | 30 | 18th | 5 | 10 | 20 | 30 |
|  | 9th | 5 | 10 | 20 | 30 | 19th | 5 | 10 | 20 | 30 |
|  | 10th | 5 | 10 | 20 | 30 | 20th | 5 | 10 | 20 | 30 |
|  |  |  |  |  |  | Deviation ($\sigma$) | 0 | 0.224 | 0.224 | 0.224 |
|  |  |  |  |  |  | $\sigma \times 100\%$ | 0 | 22.36 | 22.36 | 22.36 |
|  |  |  |  |  |  | $1 - \sigma$ | 100 | 77.64 | 77.64 | 77.64 |

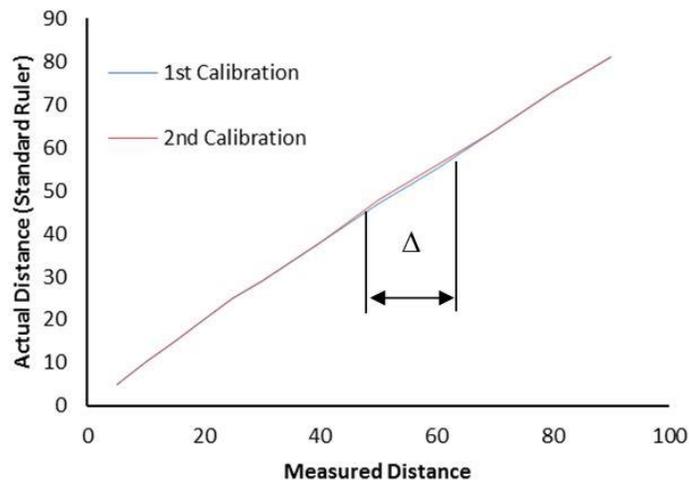

**FIGURE 3.** Two calibrating cycle to determine the repeatability error

The next parameter is repeatability error. Repeatability error ($\delta_r$) is caused by the inability of an instrument to represent the same value under presumably identical conditions. The repeatability is expressed as a maximum difference between the output as determined by two calibrating cycles ($\Delta$). It is usually represented as % of full scale ($FS$) as expressed in Eq. 5 (Fraden 2010).

$$\delta_r = \frac{\Delta}{FS} \times 100\% \qquad (5)$$

$$\delta_r = \frac{10}{90} \times 100\% = 11.11\% \qquad (6)$$

The repeatability error of this instrument prototype is 11.11%. It is obtained by comparing two calibration cycle generated in the laboratory as shown in Figure 3. As mentioned before, the nature of ultrasonic which is sensitive to change in direction contributes to the error. In the next development, by utilizing low power laser to keep the probes facing each other is able to decrease the repeatability error as well as increases the accuracy.

## Nonlinearity of Instrument

Nonlinearity error is specified for instruments whose transfer function may be approximated by straight line. A nonlinearity is a maximum deviation of a real transfer function from the straight line approximation (Fraden 2010). This research uses linear model which is derived from linear regression. The relationship between measured distance and ultrasonic wave propagation time is shown in Figure 4. Using linear regression method, PDM has $R^2$ correlation parameter equal to 99.84%, it means the model and the distance measured are correlated by 99.84%. Thus pragmatically it can be said that the nonlinearity is 1 minus the $R^2$ or 0.16%.

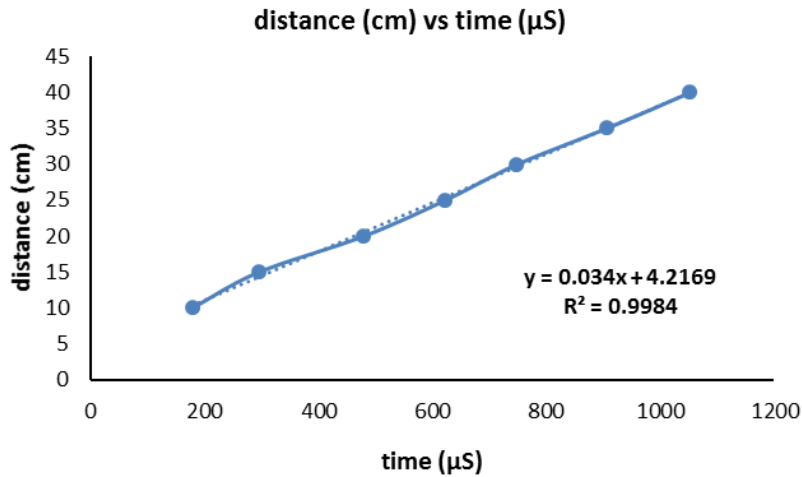

**FIGURE 3.** Relationship between measured distance and ultrasonic propagation time.

## User Interface and Storage Media

The result of the measured value is served to the user by two methods. First by LCD display which is embedded in the instrument, and second by speaker provided in the instrument. Correlated as previous discussion about resolution, by using programming, Arduino displays the measured value in unit of centimeters with resolution of 1 cm, by using an algorithm and pre-recorded mp3 clip, the output of the instrument is converted into a sequence of sounds that make up the numbers of measurement result. The instrument has a menu that is provided to measure a distance, calibrate the instrument, play the measured distance sound, and save the measurement result into storage media.

The SPI interface is utilized to embed an algorithm for saving the measurement result in Comma Separated Values (CSV) that compatible with spreadsheet program such as Microsoft Excel. Beside that, it is also compatible with database platform like MySQL which is important to the next development such as integration with internet of things. The Package of Instrument prototype is shown in Figure 5.

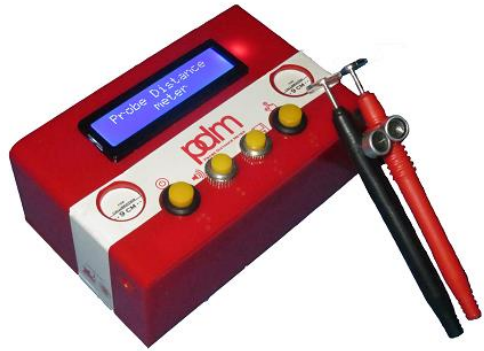

**FIGURE 4.** PDM (Probe Distance Meter) package

## CONCLUSION

PDM prototype is capable to measure properly over a range of 3 cm - 200 cm with resolution of 1 cm. Although measuring distance using ultrasonic is not new, it is proven empirically that we can measure a distance simply by marking two references with ultrasonic probes instead of using ultrasonic wave reflection. The novelty of this distance instrument is in the measurement method that is using two probes as reference points. The operation is very simple just putting a probe on the reference point and putting the other probe on the point to be measured. The measurement result is displayed on the LCD screen. The instrument has accuracy of 77.6% and small nonlinearity which is 0.16%. The laboratory test shows that the problem is not in the linearity of the instrument, but more on the repeatability error and the resolution which can be decreased by upgrading the quality of ultrasonic transducers and installing low power laser to keep the probes facing each other in one line of direction. PDM is a digital measuring instrument which has MicroSD card as storage media, thus the measurement results can be stored automatically and shown in a spreadsheet program such as Microsoft Excel. This instrument is also equipped with braille and sound module, thus it can be used by blind people who want to take distance measurement. PDM is still in development and it is potential to be a model of digital measuring devices in the future.

## ACKNOWLEDGMENTS

The authors thank the Ministry of Research and Higher Education (RISTEKDIKTI) Republic of Indonesia for giving PKM research grant for 1 year (2014-2015) and Department of Electrical Engineering and Information Technology Universitas Gadjah Mada for giving research facility and supports.